\documentstyle[times,psfig]{jaa}
\begin{document}
\title[21cm line Obs. of Interstellar Clouds using the GMRT]
{GMRT Observations of Interstellar Clouds in the 21cm line of
Atomic Hydrogen} 
\author[R. Mohan {\it et al}]%
       {Rekhesh Mohan,$^1$\thanks{e-mail:reks@rri.res.in}
        K.S. Dwarakanath,$^1$ G. Srinivasan,$^1$
\newauthor 
\& Jayaram N. Chengalur$^2$ \\
$^1$Raman Research Institute, Bangalore 560 080, India \\
$^2$National Center for Radio Astrophysics, Ganeshkhind, Pune 411 007, India \\}
\pubyear{2001}
\volume{22}
\date{Received 2000 October 23; accepted 2001 January 10.}
\maketitle
\label{firstpage}
\begin{abstract}
Nearby interstellar clouds with high ($|v| \ge 10 {\rm km~s}^{-1}$) 
random velocities although easily detected in NaI and CaII lines have 
hitherto not been detected (in emission or absorption) in the HI 21cm line.
We describe here deep  Giant Metrewave Radio Telescope (GMRT) HI absorption 
observations toward radio sources with small angular separation from
bright O and B stars whose spectra reveal the presence of intervening
high random velocity CaII absorbing clouds. In 5 out of the 14 directions
searched we detect HI 21cm absorption features from these clouds. The 
mean optical depth of these detections is $\sim 0.09$ and FWHM is 
$\sim 10~{\rm km~s}^{-1}$, consistent with absorption arising 
from CNM clouds. 
\end{abstract}


\section{Background and motivation}
\label{sec:bg}
This study is a part of our continuing effort to detect interstellar clouds in
the lines of sight to bright O and B stars in absorption in the 21cm line of
neutral hydrogen. Such clouds have been extensively studied using absorption
lines of singly ionized calcium. The most famous of such investigations is that
by Adams (1949), who used the CaII absorption spectra towards 300 stars to
determine their radial velocities. In an equally seminal paper Blaauw (1952)
used Adams' data to obtain a histogram of the random velocities of interstellar
clouds, often (roughly) allowing for the contribution to their radial velocity
due to the differential rotation of the Galaxy. One of Blaauw's main
conclusions was that interstellar clouds have significant peculiar velocities,
with the tail of the distribution extending upto 80 - 100 km s$^{-1}$. In an
independent investigation, Routly \& Spitzer (1952) found an interesting
systematic behavior of the ratio of the column density of neutral Sodium to
singly ionized Calcium (N$_{NaI}$/N$_{CaII}$). They found this ratio to decrease
with increasing random velocity of the cloud. The ratio was less than 1 for the
faster clouds ($|v|$ $\ge$ 20 km s$^{-1}$), but significantly greater than 1 in
clouds with smaller random velocities. This seems to suggest two ``classes'' of
interstellar clouds: those with negligible random velocities and those with
substantial random velocities ($|v|$ $>$ 15-20 km s$^{-1}$).

Soon after the 21cm line of the hydrogen atom was discovered, there were
attempts to detect atomic hydrogen in the interstellar clouds along the lines
of sight studied by Adams (Habing 1968,1969; Goldstein \& MacDonald 1969).
Interestingly, whereas many lines of sight towards the O and B stars studied by
Adams showed the presence of several clouds - some with very small random
velocities, while others with larger random velocities - only the low velocity
ones ($|v|$ $<$ 10 km s$^{-1}$) were seen in emission in the 21cm line. The
emission occurred at velocities that agreed well with the velocities of the CaII
absorption line features. For some reason, there was no emission that could be
attributed to the higher (random) velocity clouds. Although this was very
intriguing, after some speculations it was soon forgotten. Recently Rajagopal
{\it et al}. (1998a, 1998b) revived this question. Unlike in earlier 21cm studies,
they attempted to detect the atomic hydrogen in these clouds in a 21cm
$\it{absorption}$ study. They selected some two dozen stars from the Adams'
sample and used the VLA to do a 21cm absorption measurement against known radio
sources whose angular separation from the star in question was within a few
arc minutes. The idea was that such a line of sight to the radio source would
pass through the same interstellar clouds that were detected earlier through
the absorption lines of CaII. Surprisingly the conclusions of this absorption
study were the same as that of the earlier emission studies: Only the clouds
with random velocities less than $\sim$ 10 km s$^{-1}$ were detected (and in
each case the 21cm absorption velocity agreed with the velocity of the 21cm
emission feature and also with the velocity of the CaII absorption line). No
21cm absorption was detected from the clouds with random velocities in excess
of 10 km s$^{-1}$ down to an optical depth of 0.1. 

To explain this, Rajagopal {\it et al} (1998b) invoked the hypothesis that the
peculiar velocities of interstellar clouds was due to their encounters with
expanding supernova remnants. They argued that if this was the operative
mechanism then the clouds accelerated to higher velocities would be warmer (due
to their being dragged along by the hot gas behind the SNR shock front) and will
also have smaller column densities compared to the slower clouds (due to
evaporation). This would explain the small optical depth for the 21cm absorption
(which is directly proportional to the column density and inversely
proportional to the temperature). The Routly-Spitzer effect referred to
earlier, will also have a natural explanation in this hypothesis. In clouds
shocked by supernova blast waves, less of Calcium is likely to be locked up in
grains due to sputtering of the grains. 

Although shock acceleration of clouds seemed to provide satisfactory
explanation, to strengthen that hypothesis it was important to observationally
$\it{establish}$ a correlation between the 21cm optical depth and the peculiar
velocity. The present study is a part of a continuing effort towards this end.
In this paper, we wish to report the results obtained from the first set of
observations with the recently constructed Giant Meterwave Radio Telescope
(GMRT). In these observations we obtained a limit on the optical depth roughly
ten times better than that achieved by Rajagopal {\it et al} (1998a). The strategy for
selecting the sources is outlined in the next section and some details
pertaining to the observations are given in section 3. The results obtained by
us are given in section 4 and a brief discussion in section 5.

\section{Source selection}
\label{sec:sources}
The basic finding list of stars with the CaII absorption line data was from
Adams (1949) and Welty {\it et al}. (1996). The selection criteria was that the
spectra of the stars should contain both low ($|v|$ $<$ 10 km s$^{-1}$) and 
high random velocity ($|v|$ $>$ 10 km s$^{-1}$) optical absorption features.
Even though it is not possible to derive a sharp distinction between the low
and the high random velocities, we have adopted the value 10 km s$^{-1}$ as the
dividing line (Spitzer, 1978). The distances to the stars were obtained from
the HIPPARCOS catalog (1997). The background radio sources towards which we
measured HI absorption were selected from the National Radio Astronomy
Observatory Very Large Array Sky Survey (NVSS, Condon {\it et al.} 1996). Rajagopal
{\it et al} (1998a) chose their directions such that at half the distance to the
star, the linear separation between the lines of sight towards the star and the
radio source was $\sim$ 3 pc. In the present observations, we have chosen the
directions where this value is $<$ 1 pc in most cases. This gives a better
chance for both lines of sight to sample the same gas. Since our aim was to
reach an rms of 0.01 in optical depth, only those radio  sources with flux
densities at 20cm greater than $\sim$ 100 mJy were considered, so that we could
reach the target sensitivity in a reasonable integration time. However, for a
few directions we compromised the separation between the lines of sight for the
higher flux density of the background source. Our final list consisted of 14
fields. Table~1 lists a summary of the  fields observed. In a few cases, there
was more than one radio source within the GMRT primary beam. Four out of the
fourteen stars in our list have been previously studied by Rajagopal
{\it et al} (1998) using the VLA.

\begin{table}[h]
\caption{A summary of the observed directions: Column 1 lists the stars, whose
spectra are known to contain high random velocity optical absorption lines.
Those fields which were observed earlier by Rajagopal et. al (1998a) is marked
with an asterix ($^*$). Column
2 shows the distance to these stars obtained form the HIPPARCOS catalog. The
flux density of the radio sources listed in column 6 are from the present
observations. The angle $\theta$ (column 7) is the angle between the lines of
sight to the star and the radio source. The linear separation between the two
lines of sight at half the distance to the star is given in column 8} 
\label{1}
\begin{center}
\begin{tabular}{ l c c c c c c c r } \hline \hline
Star       & d   &l     & b      &Radio source& S & $\theta$  & r    \\
            & pc  &deg   & deg    &            & mJy     & & pc   \\ \hline
HD 175754$^*$& 680.0 & 16.46 & $-$10.02 & NVSS J1856$-$192& 67 & 28$^{'}$ & 2.8 \\
HD 159561 & 14.3  & 35.90  & +22.58 & NVSS J1732+125 &174& 41$^{'}$  & 0.1 \\
HD 166182 &467.3 & 47.42  & +18.03 & NVSS J1809+208 &383& 13$^{'}$  & 0.9 \\
HD 193322 &476.2 & 78.10  & +2.78  & NVSS J2019+403 &310& 28$^{'}$  & 1.9 \\
HD 199478$^*$&2857  & 87.51 & +1.42  & NVSS J2056+475 &180& 8$^{'}$  & 3.3 \\
HD 21278$^*$&174.8 & 147.52  & $-$6.19 & NVSS J0330+489 &343& 28$^{'}$  & 0.7 \\
          &      &         &       & NVSS J0331+489 &189& 22$^{'}$  & 0.6 \\
HD 24760  & 165.0  & 157.35 & $-$10.10 & NVSS J0400+400&191& 30$^{'}$  & 0.7 \\
HD 47839  & 313.5  & 203.00 & +2.30 & NVSS J0642+098 &340& 8$^{'}$  & 0.4 \\
HD 37128  & 411.5 & 205.26  & $-$17.14  & NVSS J0536$-$014 &171& 18$^{'}$  & 1.0 \\
HD 37742$^*$&250.6  & 206.50 & $-$16.49 & NVSS J0542$-$019&408 & 7$^{'}$  & 0.3 \\
HD 37043 & 406.5 & 209.50  & $-$19.60 & NVSS J0535$-$057&299 & 14$^{'}$  & 0.9 \\
         &       &         &        & NVSS J0536$-$054&215 & 21$^{'}$  & 1.2\\
HD 38771 &221.2  & 214.52 & $-$18.50 & NVSS J0549$-$092 &170& 32$^{'}$  & 1.0 \\
           &       &         &       & NVSS J0549$-$092 &97 & 32$^{'}$  & 1.0 \\
HD 143018 &140.8 & 347.20 & +20.14 & NVSS J1559$-$262 &328& 10$^{'}$  & 0.2 \\
HD 147165 &225.2 & 351.38 & +16.90 & NVSS J1623$-$261 &287& 33$^{'}$  & 1.3 \\ \hline \hline
\end{tabular}
\end{center}
\end{table}

\section{Observations}
\label{sec:obs}
\subsection{The Giant Meterwave Radio Telescope}  
The Giant Meterwave Radio Telescope (GMRT) consists of 30 fully steerable
dishes, of diameter 45 m with a maximum baseline of 25 km (Swarup {\it et al},
1991). The aperture efficiency of the dishes is $\sim$ 40$\%$ in the 21cm band,
which implies an effective area of $\sim$ 19000 m$^{2}$. This telescope is
equipped with an FX correlator providing 128 channels per polarization per
baseline. A baseband bandwidth ranging from 16 MHz down to 64 kHz variable in
steps of 2 can be  chosen. The 21 cm receiver is a wide band system covering
the frequency range of 900 to 1450 MHz. It is a prime focus uncooled receiver
with a characteristic system temperature of 70~K. The 21cm system has four sub
bands, centered at 1060, 1170, 1280 and 1390~MHz, each with a 3 dB bandwidth of
120 MHz. Provision exists in  the receiver to bypass the narrow bandpass
filters to obtain the full 450 MHz  band.

\subsection{Observing Strategy} 
The radio observations were carried out during the period April-May, and
September, 1999. Only 8-10 antennas within the central square km of the array
were used. We used a baseband bandwidth of 2 MHz,  which translates to 
$\approx$ 422 km s$^{-1}$ in velocity and a resolution of $\sim$ 3.3 km
s$^{-1}$. The center of the band was set at 1420.4 MHz. The
observing band was found to be free from interference. We used one of the VLA
primary flux calibrators (3C48/3C147/3C286) for setting the flux density scale.
We observed a nearby secondary calibrator from the VLA calibrator manual for
phase and bandpass calibration. The phase calibration was carried out once
every hour. Bandpass calibration was carried out once every two hours for 10
minutes with the observing frequency shifted by 2 MHz below the band, which
corresponds to a velocity shift of $\approx$ 400 km s$^{-1}$. This shift is
adequate to move the observing band out of the velocity spread of the Galactic
21cm line spectral features. On source integration time ranged from $\sim$ 1 to
7 hours, depending on the strength of the  background radio source. The rms
sensitivity in optical depth varied from 0.02 to 0.007 towards
different sources. A summary of the observational setup is given in Table~2. 

The analysis of the data was carried out using the Astronomical Image
Processing System (AIPS) developed by the National Radio Astronomy Observatory.
The resulting data set consisted of 14 image cubes containing a total of 17 HI
absorption profiles. Continuum subtraction was carried out by fitting a linear
baseline to the line free channels in the visibility domain and subtracting the
best fit continuum from all the channels. For the point sources amongst the
list of program sources, the flux densities quoted in the NVSS (Condon {\it et al},
1996) was found to agree with the flux densities obtained from the GMRT to
within 10\%. Separate line  images of short and long  baselines were made to
convince ourselves that contamination due to HI emission was minimal.  The rms
noise level ranged from 1.5 mJy to 5 mJy/channel/beam depending upon the
integration time. Out of the 17 directions, the lowest HI optical depth of
$\sim$ 0.007 was achieved towards the source NVSS J0542$-$019. In order to study
the individual HI absorption components, multiple Gaussian profiles were fitted
to the absorption line spectra using the Groningen Image Processing System
(GIPSY). 

\begin{table}
\caption{The Observational Setup.}
\label{2}
\begin{center}
\begin{tabular}{ r l } \hline \hline
Telescope & GMRT \\
System temperature & $\sim$ 70 K \\
Aperture efficiency & $\sim$ 40$\%$ \\
Number of antennas & 8 to 10 \\
Base band used & 2.0 MHz \\
Number of channels & 128 \\
Velocity resolution& 3.3 km s$^{-1}$ \\
On source integration time & $\sim$ 1 to 7 hour \\ \hline \hline
\end{tabular}
\end{center}
\end{table}

\section{Results}
\label{sec:results}
We have examined the HI absorption spectra towards 14 stars having optical
absorption lines at both low and high random velocities. In all these fields,
we have detected HI absorption at low random velocities. In all the directions,
at low random velocities ($|v|$ $<$ 10 km s$^{-1}$) there is a good agreement
between the velocities of the HI absorption features and that of the CaII 
absorption line components, within the limits of our velocity resolution 
($\sim$ 3.3 km s$^{-1}$). In 5 out of the 14 fields, for the first time, we
have detected HI absorption features coincident with the high random velocity 
CaII absorption lines. 

The 21cm absorption profiles have a typical optical depth 
$\tau \sim 0.1$, and a FWHM $\sim 10$~km~s$^{-1}$. This velocity width
is less than would be expected from thermal broadening in the WNM, and
in any case, the path length of WNM required to produce an optical 
depth of $\sim 0.1$ is considerably larger than the distance to the
star against which the NaI/CaII lines have been observed. The observed
velocity width and optical depth are instead completely consistent with
what would be expected from absorption in a CNM cloud. In contrast with
the classical galactic emission/absorption line studies however, this
cloud is too small to produce a clearly identifiable single dish 
emission signal. An estimate of the column density can however be 
obtained from the Leiden-Dwingeloo Survey (LDS) of Galactic neutral 
hydrogen (Hartmann \& Burton, 1995). Note that the column density as
measured in this survey is at best indicative, since beam dilution causes
it to underestimate the true N$_{HI}$ of the cloud while contamination
from the WNM causes it to overestimate the true $N_{HI}$. Nonetheless
we have used the LDS HI column density to get a handle on the spin 
temperatures of the  high random velocity clouds that we have detected. 
The N$_{HI}$ was estimated by integrating over the FWHM of the HI 
absorption line. The same procedure was repeated for the low LSR 
velocity HI absorption features in the spectra and gave values 
that were consistent with earlier estimates. However, note that
for the smaller clouds at higher velocity  the fractional 
contribution from the WNM would  be more in the HI emission spectra 
and would hence lead to greater errors in the estimation of 
N$_{HI}$ and thus T$_S$. 

Each of the five detections are discussed separately below. A summary of the
detections is listed in Table~3. Since our velocity resolution is $\sim$ 3 km
s$^{-1}$ and the optical absorption data for majority of the stars were
obtained from a high resolution (0.3 - 1.2 km s$^{-1}$)  survey (Welty {\it et al}
1996), the number of interstellar absorption features seen in the optical study
is greater than  those revealed by our HI absorption study. This is evident
in figures 1 to 6 which display optical depth profiles. 

\begin{table}
\caption{Detections of high random velocity HI absorption: 
Column 2 lists the high velocity ($|v|$ $>$ 10 km s$^{-1}$) CaII absorption
lines seen towards the star. Column 3 gives the
radial component of the Galactic rotation velocity at the distance to the 
star. Column 4 gives the HI optical depth at the velocity of the CaII 
absorption line
and Columns 5,6 and 7 list the parameters of the gaussian fit to the HI
absorption profile. Column 8 lists the HI column density in the respective
line of sight derived from the Leiden Dwingeloo Survey (Hartmann \& Burton,
1995) and Column 9 gives the derived HI spin temperature.}
\label{3}
\begin{center}
{\small{
\begin{tabular}{ l c c c c c c c r } \hline \hline
Star& CaII      & v$_{Gal}$ &   & Mean & Peak   & FWHM & N$_{HI}$  & T$_{S}$ \\
HD no.    &  V$_{lsr}$  &       & $\tau_{HI}$ & V$_{lsr}$ & $\tau_{HI}$   &    & $\times$10$^{20}$ &  \\
    & (kms$^{-1}$)         &(kms$^{-1}$)&  & (km s$^{-1}$) &      & (km s$^{-1}$) & (cm$^{-2}$) & (K) \\ \hline
37043 &$-$16.86  &   +2.5  &  0.12 & $-$15.1 & 0.12 & 12.8 & {{0.2}$^{\dagger}$}  & 7$^{\dagger}$ \\
         &$-$12.34  &         &  0.10 &       &      &      &   & \\
         &+22.68  &         &  0.06 & +24.0 & 0.06 & 20.4 & 1.8  & 78 \\
159561&$-$14.07  &  +1.9   &  0.06 & $-$13.0 & 0.06 & 11.6 & 0.5  & 40 \\
193322&+21.68  &  +5.2   &  0.36 & +24.0 & 0.24 & 6.3  & 2.6 & 89 \\
143018&$-$26.48  &  $-$1.4   &  0.04 &       &      &      &   & \\
         &$-$21.63  &         &  0.07 &       &      &      &   & \\
24760 &+12.96  &$\sim$ 0 &  0.06 &       &      &      &   & \\ \hline \hline
\end{tabular}
}}
\end{center}
\noindent{{\footnotesize {{$^{\dagger}$: Upper limit, possibility of beam dilution.}}}}
\end{table}

\subsection{HI absorption detections from high random velocity clouds.}
\noindent{\bf{HD37043/NVSS J0535$-$057:}} Towards the star HD37043, the  high
resolution CaII absorption line study by Welty i{\it et al} (1996) revealed 12
interstellar absorption features over the velocity range $-$22 to +23 km
s$^{-1}$.   The HI absorption was measured towards {\mbox{NVSS J0535$-$057}}, 
which is 14$^{'}$ in projection from the star. This angular separation is
equivalent to a linear separation of 0.9 pc at half the distance to the star.
The HI optical  depth profile shows weak but significant absorption 
corresponding to the high random velocity CaII absorption lines (Figure~1). 
This weak HI absorption persists over the velocity spread of the
CaII absorption lines. The HI absorption profile was fitted with 5 gaussians,
centered at +24, +12, +5, $-$2 and $-$15km s$^{-1}$. For the direction towards this
star, negative velocities are forbidden by the Galactic rotation model (Brand
\& Blitz, 1993). Presumably due to the lack of adequate velocity resolution
in the 21cm spectra, the individual HI absorption features are  spread 
over many optical absorption line components and it is difficult to obtain 
a one to one correspondence between them. 

\begin{figure}
\psfig{file=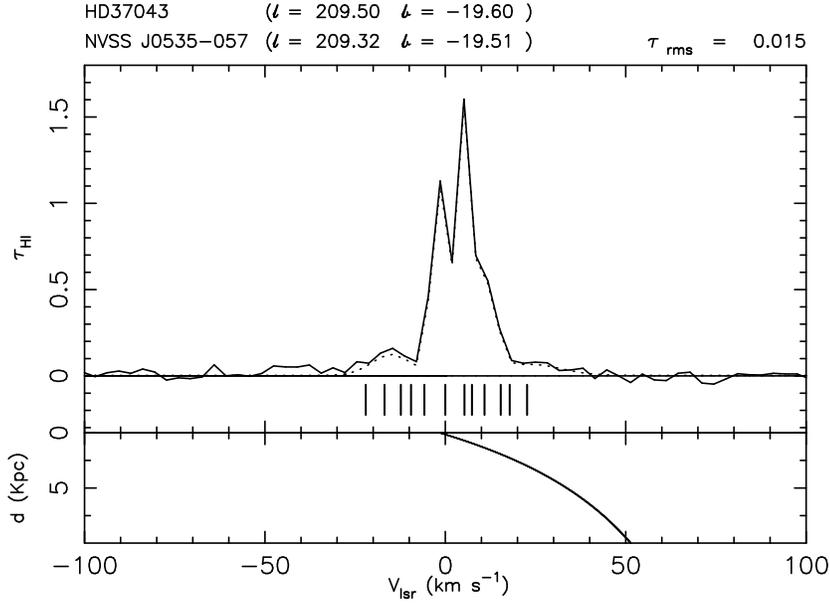,width=12.0cm,angle=-90}
\caption{The HI optical depth spectrum (solid line) towards the radio
source NVSS J0535$-$057. The vertical tick marks indicate the velocities of the
CaII absorption lines in the spectrum towards the star HD37043. The radio
source and the star are separated by 14$^{'}$ in the sky, which is equivalent
to a linear separation of $\sim$ 0.9 pc at half the distance to the star. The
weak HI absorption extends over the velocity spread of the CaII absorption
lines, with a distinct  absorption feature at v$_{lsr}$ $\sim$ $-$15 km s$^{-1}$.  The
broken curve plotted along with the optical depth profile is the model from the
Gaussian fitting to the absorption spectrum. The lower panel shows the radial
component of Galactic rotational velocity for this direction as a function of
heliocentric distance.}
\label{1}
\end{figure}

The HI absorption feature centered at around $-$15 km s$^{-1}$ coincides with the
optical absorption line at $-$16.86 km s$^{-1}$, well within the limits of our
velocity resolution. The FWHM of this HI feature is $\approx$ 13 km s$^{-1}$
and the peak optical depth is $\approx$ 0.12. In the LDS survey, there is 
no distinct HI emission feature at the velocity of the CaII and HI absorption
lines, but only a shoulder of HI emission.If the gas associated with this
cloud fills the Dwingeloo telescope beam, then the LDS provides an upper 
limit of  2.0 $\times$ 10$^{19}$ cm$^{-2}$ for the HI column density of this
cloud. The upper limit on HI spin temperature implied by the N$_{HI}$ 
from LDS is 7~K, which is about an order of magnitude lower than typical 
diffuse interstellar HI clouds. It seems highly likely therefore that
the high random velocity optical and HI absorption arises from a cloud 
much smaller in size compared to the Dwingeloo telescope beam. 

Corresponding to the CaII line at +22.68 km s$^{-1}$, the HI absorption feature
at +24 km s$^{-1}$ has a derived N$_{HI}$ $\sim$ 1.8 $\times$ 10$^{20}$
cm$^{-2}$ from the LDS. The implied spin temperature for this cloud is $\sim$
78 K. 

The second radio source NVSS J0536$-$054 in the same field of view is at an
angular distance of 21$^{'}$ from the star (corresponding to 1.2~pc at half 
the distance to the star). Interestingly, even though HI absorption is 
seen over a wide range of velocities towards this source, the correlation 
of CaII and HI absorption line positions at higher velocities is poor 
(Figure~2). In fact, the high velocity feature at $\sim$ $-$15 km s$^{-1}$ 
seen clearly in Fig. 1 is absent in this figure. Obviously, this line of sight
is outside the HI absorbing cloud. 

\begin{figure}
\psfig{file=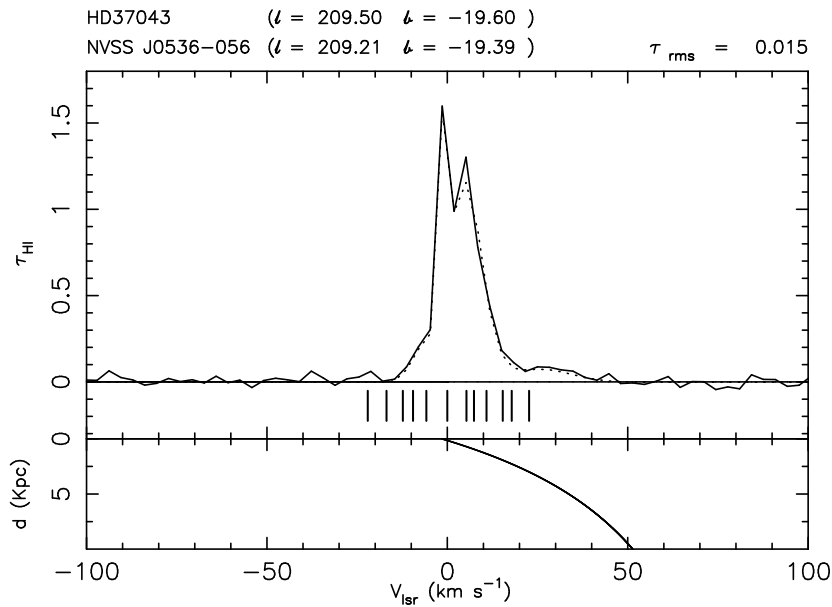,width=12.0cm,angle=-90}
\caption{The HI optical depth profile towards the radio source NVSS J0536$-$054.
The radio source is $\sim$ 21$^{'}$ away  in projection from the star HD37043. 
which
implies a linear separation of $\sim$ 1.2 pc at half the distance to the star.
The high random velocity HI absorption feature at $\sim$ $-$15 km s$^{-1}$ seen
towards the radio source NVSS J0535$-$057 (fig 1) is not detected here.} 
\label{2}
\end{figure}

\noindent{\bf{HD159561/NVSS J1732+125}:} The CaII absorption line data obtained
from Welty {\it et al} (1996)  indicates three interstellar CaII absorption features,
one of which is at a high random  velocity of $\approx$ $-$14 km s$^{-1}$ (Figure
3). The background source for the HI absorption study was NVSS J1732+125,
41$^{'}$ in projection from the star. This angular extent is equivalent to a
linear separation of 0.1 pc at half the distance to the star. Multiple 
gaussian fitting to the HI  absorption profile provided 4 components,
centered at +24, +12, $-$0.5
and  $-$13 km s$^{-1}$. The feature at $-$13 km s$^{-1}$, with FWHM of 12 km
s$^{-1}$ and a peak optical depth of $\sim$ 0.06 coincides with the CaII
absorption line at $-$14 km s$^{-1}$. At a longitude of 36$^{o}$, this velocity
is forbidden by the model of Galactic rotation (Brand \& Blitz 1993). Given a
distance to the star  $\sim$ 14 pc, CaII and the HI absorption should be from a
cloud in the local neighborhood. The HI column density estimated from the 
LDS is $\sim$ 5.3 $\times$ 10$^{19}$ cm$^{-2}$. This implies a spin 
temperature  $\sim$ 40 K.

\begin{figure}
\psfig{file=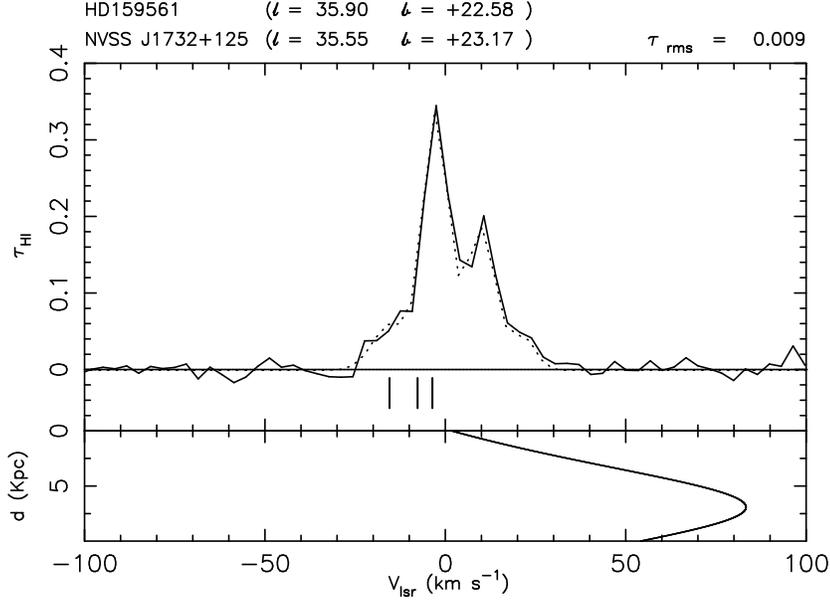,width=12.0cm,angle=-90}
\caption{The HI optical depth profile towards the radio source NVSS J1732+125.
The vertical tick marks indicate the velocities of the CaII absorption lines in
the spectrum towards the star HD159561. The line of sight towards the radio
source is separated by $\sim$ 0.1 pc from that towards the star  at half the
distance to the star. The HI absorption feature at $\sim$ $-$13 km s$^{-1}$
coincides well with the CaII absorption line. The dotted curve plotted along 
with the optical depth profile is the model from the gaussian fitting to the
absorption spectrum. The lower panel shows the radial component of Galactic
rotational velocity in this direction as a function of heliocentric distance.} 
\label{3}
\end{figure}

\noindent{\bf{HD193322/NVSS J2019+403}:} The star HD193322 was observed by
Adams (1949) for interstellar CaII absorption lines.  This direction is at a
low Galactic latitude (b $\sim$ 2$^{o}$) and hence the  HI absorption profile
towards an extra-galactic radio source can be complicated with a large number
of absorption features arising from cold clouds in front of the star as well as
from behind it. However it was chosen  for our HI absorption study due to the
presence of CaII absorption lines at velocities forbidden by the Galactic
rotation.  The HI absorption was measured towards NVSS J2019+403 which is 
28$^{'}$ in projection from the star. This angular separation is equivalent  to
a linear separation of about 1.9 pc at half the distance to the star. The HI
optical depth profile is complicated as expected (Figure 4). The absorption
profile was fitted with 9 gaussian components. There is a prominent component
at an LSR velocity of $\approx$ +24 km s$^{-1}$ near the CaII absorption
component at +21.68 km s$^{-1}$. The FWHM of this HI  absorption feature is
$\sim$ 6 km s$^{-1}$ and the peak optical depth is $\sim$ 0.24. The N$_{HI}$
obtained from the LDS is $\sim$ 2.6 $\times$ 10$^{20}$ cm$^{-2}$, and the HI
spin temperature $\sim$ 89 K. However, we failed to detect HI absorption at the
velocity of the CaII absorption line at +31 km s$^{-1}$ (Fig 4).

\begin{figure}
\psfig{file=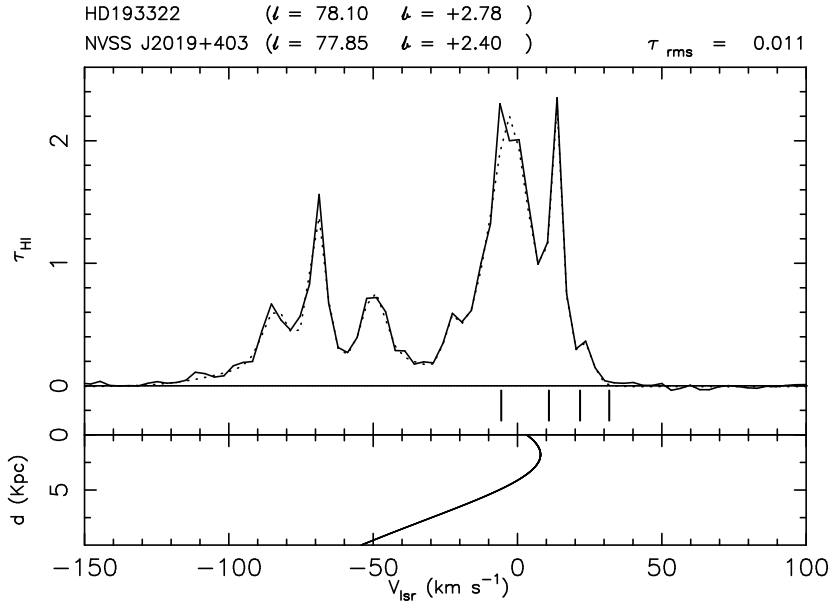,width=12.0cm,angle=-90}
\caption{The HI optical depth profile towards the radio source NVSS J2019+403.
The vertical tick marks indicate the velocities of the CaII absorption lines in
the spectrum towards the star HD193322, which is at an angular separation of
$\sim$ 28$^{'}$ from the radio source in the sky. The line of sight towards the
radio source is separated by $\sim$ 1.9 pc from that towards the star  at half
the distance to the star. The CaII absorption line at $\sim$ 23 km s$^{-1}$ has
a coincident HI absorption feature. The dotted curve plotted along with  the
optical depth profile is the model from the gaussian fitting to the absorption
spectrum. The lower panel shows the radial component of Galactic rotational
velocity in this direction as a function of heliocentric distance.} 
\label{4}
\end{figure}

\noindent{\bf{HD143018/NVSS J1559$-$262}:} The CaII absorption data for this star
is from Welty {\it et al} (1996). The interstellar CaII absorption lines exist at
negative LSR velocities upto $-$26 km s$^{-1}$. Since the star is located at a
distance of 141 pc,   the radial component of Galactic rotational velocity is
only  $-$1.4 km s$^{-1}$ at the distance of the star. Hence, the CaII lines are
arising from high random velocity clouds in the interstellar space. The HI
absorption was measured towards NVSS J1559$-$262, 10$^{'}$ in projection from the
star, which implies a linear separation of 0.2 pc at half the distance to the
star. An HI absorption measurement towards an extragalactic source would sample
gas all along the line of sight, even beyond the star. The HI optical depth
profile is shown in Figure 5. It is clear from the figure that

\begin{itemize}
\item{There is non-zero HI optical depth at the positions of the high random
velocity CaII absorption lines.}
\item{The HI absorption decreases to zero beyond the highest velocity optical
absorption line component.}
\item{For the latitude $\sim$ +20$^{o}$, any low random velocity cloud 
(i.e. a cloud whose radial velocity is mainly due to Galactic rotation)
with an LSR velocity of  $\sim$ $-$21 km s$^{-1}$  should be at a distance 
of about 1~kpc above the Galactic plane, which places it in the Galactic 
halo. This appears unlikely. It is more likely that this HI cloud is 
in front  of the star and has a high random velocity.}
\end{itemize}

The Gaussian profile fitting provided three components, at velocities $-$12.5, $-$7
and $-$4.5 km s$^{-1}$, with 5, 32 and 3.5 km s$^{-1}$  respectively as the FWHM.
The existence of non zero HI optical depth at the  velocities of the CaII
absorption lines makes this a detection of high random velocity clouds in HI
absorption but the poor signal to noise ratio prevented a faithful Gaussian
fit to this data. However, the non zero HI optical depth at the positive LSR
velocities, extending beyond +20 km s$^{-1}$ demands more attention.

\begin{figure}
\psfig{file=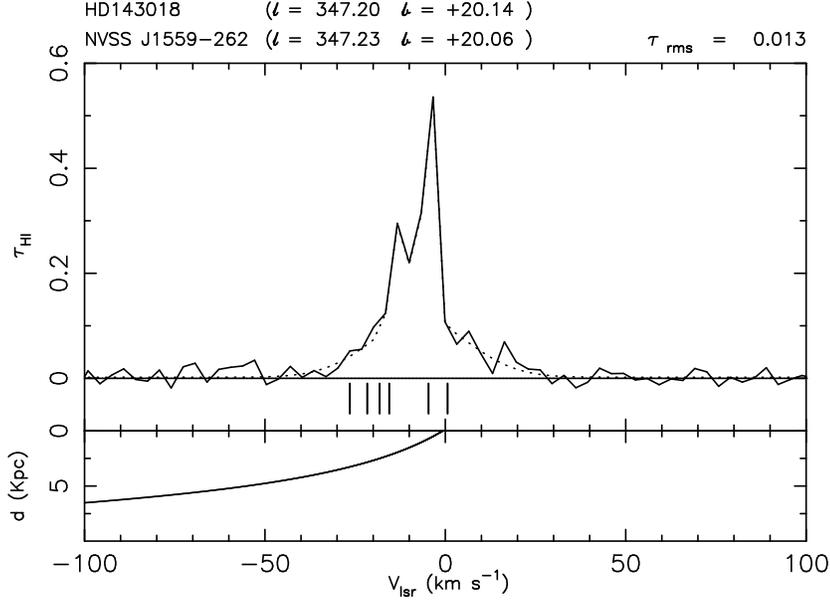,width=12.0cm,angle=-90}
\caption{The HI optical depth profile towards the source J1559$-$262 at a
distance of $\sim$ 10$^{'}$ in projection from the star HD143018 The vertical
tick marks indicate the velocities of the CaII absorption lines in the spectrum
towards the star. The lower panel shows the radial component of Galactic
rotational velocity for the direction as a function of heliocentric distance. 
For the latitude of this direction, any low random velocity cloud at $\sim$ 
$-$21 km s$^{-1}$ should be at a distance of about 1 kpc above the Galactic
plane, in the Galactic halo. Hence, any HI absorption at these velocities has
to be from a nearer cloud with a large random velocity. The dotted curve
plotted along with the optical depth profile is the model from the Gaussian
fitting to the absorption spectra.} 
\label{5}
\end{figure}

\noindent{\bf{HD24760/NVSS J1400+400:}}  The CaII absorption data was obtained from
Welty et al (1996). There are 8  discrete interstellar absorption features seen
in the line of CaII, out of which two at LSR velocities of +13 and +16.5 km
s$^{-1}$ are at high random velocities. These velocities are forbidden by the
Galactic rotation model (Brand \& Blitz 1993).  The HI absorption towards the
radio source NVSS J1400+400, which is 30$^{'}$ in projection from the star
HD24760 is shown in Figure~6.  The HI optical depth profile indicate a non zero
HI optical depth at the position of the CaII absorption line at +13 km
s$^{-1}$,  at a level of 0.06. However, as in the previous case, the gaussian
fitting to this data gave  no discrete component to the high velocity feature
(See Table~3). 

\begin{figure}
\psfig{file=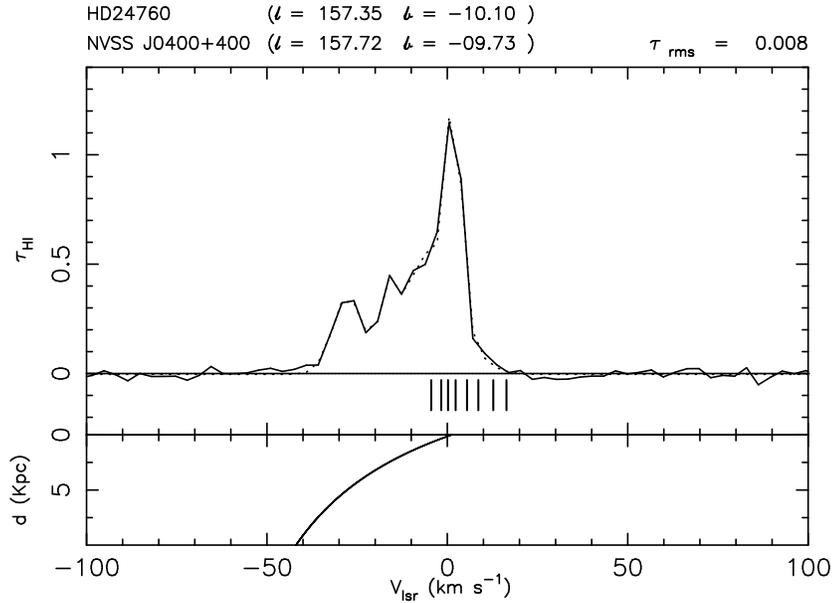,width=12.0cm,angle=-90}
\caption{The HI optical depth profile towards the source NVSS J1400+400. This
radio source is separated by $\sim$ 30$^{'}$ in projection from the star
HD24760. The vertical tick marks indicate the velocities of the
CaII absorption lines in the spectrum towards the star. There is a nonzero 
optical depth at the position of the CaII absorption
line at $\sim$ +13 km s$^{-1}$. However the gaussian fitting to the HI
absorption data gave
no discrete component to this high velocity feature. The dotted curve plotted 
along with the optical depth profile shows the model from
the Gaussian fitting to the absorption spectra.}
\label{6}
\end{figure}

\section{Summary and discussion}
\label{sec:summary}
\begin{itemize}

\item{We have obtained absorption spectra in the 21cm line of hydrogen in 14
directions which are close to the lines of sight of known bright stars against
which optical absorption studies had been done earlier.}

\item{We achieved an optical depth limit of approximately 0.01, which is about
ten times better than the previous attempts in this context.}

\item{The angular separation between the line of sight to the star and the line
of sight to the radio source is also roughly two times smaller than in the
earlier study.}

\item{In 5 out of the 14 directions we have detected HI
absorption at (random) velocities in excess of 10 km s$^{-1}$. To recall, in
their attempt Rajagopal {\it et al} (1998a) did not detect absorption from these
faster clouds.} 

\end{itemize}

As mentioned earlier, our selection criteria restricted background radio
sources to those with an angular separation from the star 
less than $\sim$ 40$^{'}$  (implying a linear separation less than $\sim$ 1 pc
at half the distance to  the star). In the previous 21cm
absorption search (Rajagopal {\it et al}., 1998a) larger angular separations
(implying linear separations $\sim$ 3 pc) were included in the sample. 
Our survey is also considerably
more sensitive  than that of Rajagopal {\it et al} (1998a). Four out of the 14 directions observed in this study - 
the fields containing the stars HD 175754, HD 199478, HD 21278 and HD 37742
respectively - were previously studied by Rajagopal {\it et
al.} (1998a, 1998b). The rms optical depth levels achieved in their study in these direction were 0.09, 0.21, 0.08
and 0.03, respectively. Whereas in our study, these figures are 0.03, 0.015,
0.008 and 0.007, respectively. However, we too failed to detect any HI 
absorption from the clouds in all these four directions. This is 
intriguing since the results of our
study indicate that about 30\% of the high random velocity clouds are
detectable (5 detections out of the 14 observed directions) at the sensitivity
levels reached in this study ($\tau_{HI}$ $\sim$ 0.01). 
For more than 50\% of the
directions studied by Rajagopal {\it et al.} (1998a, 1998b), the rms optical depth level was 
$\it{above}$
0.1. The mean value of peak optical depth for our detections is $\sim$ 0.09.
Only one of the fields studied by Rajagopal {\it et al} (1998a) had the sensitivity level to
detect this optical depth. 

Earlier we recalled the conclusion by Blaauw (1952) $viz.$ the distribution of
random velocities of clouds extended beyond $\sim$ 80 km s$^{-1}$. But this
high velocity tail seemed distinct from the rest of the distribution that was
well fit by a gaussian with a dispersion of $\sim$ 5 km s$^{-1}$. Indeed,
Blaauw was careful to remark on this. Later 21cm surveys of the Galaxy also
yielded, as a byproduct, information about the velocities of the interstellar
clouds. Unlike in the optical absorption studies towards relatively nearby
stars (with good distance estimates), in the radio observation one did not have
reliable distance estimates to the clouds and therefore one could not correct
for the Galactic differential rotation to derive a distribution of random
velocities. This was remedied by Radhakrishnan \& Sarma (1980) who analyzed the
HI absorption spectrum towards the Galactic center, for in this particular
direction the observed radial velocities of clouds could be unambiguously
attributed to random or peculiar velocities. Their analysis clearly showed that
the distribution of random velocities could be well fit with a gaussian with a
dispersion of 5 km s$^{-1}$, in excellent agreement with Blaauw's conclusion.
Radhakrishnan \& Srinivasan (1980) while interpreting the analysis of
Radhakrishnan \& Sarma (1980) suggested that whereas the narrow gaussian may represent
the population of strongly absorbing clouds, there is possible evidence for a
second population of weakly absorbing clouds, with a much larger velocity
dispersion $\sim$ 35 km s$^{-1}$. The peak optical depth of this distinct
population (if at all one could decompose them into two population) was 0.3.
Radhakrishnan \& Srinivasan (1980) interpreted these high velocity, low optical depth
clouds as those that have been shocked by supernova remnants. The existence of
a high velocity tail in the Galactic center observation has however, remained
controversial (See, for example Schwarz {\it et al} 1982 and Kulkarni \& Fitch 1985).
Not withstanding this, we wish to assume that Radhakrishnan and Srinivasan (1980) were
correct in their inference and ask if the results reported in this paper are
consistent with their findings. Let us elaborate on this part. Given the
statistics of clouds (the number of clouds per Kpc), and the long path length
to the Galactic center, the profile of 21cm optical depth as a function of
velocity can straightaway be interpreted as a velocity distribution of clouds.
In the present observations, on the other hand, one encounters only a couple of
clouds in any given line of sight. But it is worth asking if the optical depth
of clouds we have detected with particular velocities is consistent with the
optical depth versus velocity profile in the Galactic center observation.

The mean optical depth of our detections is $\tau$ $\sim$ 0.09, and the
velocities ( $|v|$ in the LSR) of the absorption features fall in the range 13
to 26 km s$^{-1}$. The distances to the stars towards which these detections
were made are between 14 to 476 pc, implying a mean distance to the absorbing
clouds $\ge$ 250 pc. The HI optical depth of the wide component discussed by
Radhakrishnan and Srinivasan was $\sim$ 0.25 in the velocity range referred
above. This value is, of course, for a path length of $\sim$ 8 Kpc to the
Galactic center, and would translate to an equivalent optical depth of $\sim$
0.01 for a distance $\sim$ 250 pc. Such an estimate may be unwarranted in view
of the uncertainities in the statistics of clouds. We merely wish to remark that
the detections reported in this paper are not inconsistent with the existing
data towards the Galactic center. As for the spin temperatures of the clouds we
have detected in absorption, a mean value $\sim$ 70 K would suggest that there
isn't much difference between the slower and faster clouds at least as far as
their spin temperatures are concerned. The tentative conclusion one can draw at
this stage is that the much smaller optical depths of the faster clouds might be
solely due to their smaller column densities.

\noindent{{\bf{Acknowledgements:}} We wish to convey our deep gratitude to the
large team of people associated with the National Center for Radio Astrophysics
in Pune for their sustained and dedicated efforts which resulted in the Giant
Meterwave Radio Telescope. We also wish to thank N.V.G. Sarma and his colleagues
at the Raman Research Institute, Bangalore for designing and constructing the
broad-band 21cm feeds for the GMRT.}

\newpage
\label{lastpage}
\end{document}